\documentclass[groupedaddress,aps,pra,superscriptaddress,showpacs]{revtex4}%
\usepackage{epsfig,dsfont,amssymb,amsmath,amsthm,amsfonts,amsbsy,mathrsfs}
\usepackage{graphicx}
\usepackage{amsmath}
\usepackage{amssymb}
\usepackage{mathdots}
\usepackage{float}
\usepackage{cases}
\usepackage{enumerate}
\begin{document}
\baselineskip18pt
\title{Geometric Global Quantum Discord of Two-qubit States}
\author{Yunlong Xiao}
\affiliation{School of Mathematical Sciences, South China University of Technology, Guangzhou 510640, China}
\affiliation{Max Planck Institute for Mathematics in the Sciences, 04103 Leipzig, Germany}
\author{Tao Li}
\thanks{Corresponding author. E-mail:~lt881122@sina.com}
\email{lt881122@sina.com}
\affiliation{School of Mathematical Sciences,  Capital Normal University, Beijing 100048, China}
\author{Shao-Ming Fei}
\affiliation{School of Mathematical Sciences, Capital Normal University, Beijing 100048, China}
\affiliation{Max Planck Institute for Mathematics in the Sciences, 04103 Leipzig, Germany}
\author{Naihuan Jing}
\affiliation{School of Mathematical Sciences, South China University of Technology, Guangzhou 510640, China}
\affiliation{Department of Mathematics, North Carolina State University, Raleigh, NC 27695, USA}
\author{Xianqing Li-Jost}
\affiliation{Max Planck Institute for Mathematics in the Sciences, 04103 Leipzig, Germany}
\author{Zhi-Xi Wang}
\affiliation{School of Mathematical Sciences,  Capital Normal University,  Beijing 100048, China}

\begin{abstract}
We consider the geometric global quantum discord (GGQD) of two-qubit systems. By analyzing the symmetry of geometric global quantum discord we give an approach for deriving analytical formulae of the extremum problem which lies at the core of computing the GGQD for arbitrary two-qubit states. Furthermore, formulae of GGQD of arbitrary two-qubit states and some concrete examples are presented.

\end{abstract}
\maketitle

\section{Introduction}
The quantum correlations\cite{1} between the subsystems $\mathfrak{A}$ and $\mathfrak{B}$ of a bipartite system play significant roles in many information processing tasks\cite{2} and applictions\cite{3,4,5}. It can be classified according to the probability distributions of the measurement outcomes from measuring the subsystems $\mathfrak{A}$ and $\mathfrak{B}$. For any quantum entangled states, the probability distributions of the measurement outcomes from measuring the subsystem $\mathfrak{A}$ will depend on the probability distributions of the measurement outcomes from measuring the subsystem $\mathfrak{B}$. Nevertheless, it is still possible that the correlations between the measurement outcomes from measuring the subsystem $\mathfrak{A}$ and from measuring the subsystem $\mathfrak{B}$ can be described by classical probability distributions. A quantum state is pronounced to hold a local hidden variable model (LHV) if all the measurement results can be modeled as a classical random distribution over a probability space. The states admitting LHV models do not violate any Bell inequalities, while the states that do not admit any LHV models violate at least one Bell inequality\cite{6,7,8}.

For separable states, the probability distributions of measurement outcomes from measuring the subsystem $\mathfrak{A}$ are independent of the probability distributions of the measurement outcomes from measuring subsystem $\mathfrak{B}$. However, these separable states may be further classified as classically correlated states and quantum correlated ones, depending on the possibility of memorizing all the mutual information by evaluating one of the subsystems. Such property is characterized by so called quantum discord\cite{9,10,11,12}. It has been shown that the quantum discord is required for some information processing like assisted optimal state discrimination \cite{13,14}.

In recent years more relevant measures such as geometric quantum discord\cite{15,16,17} (GQD) have been suggested. It takes use of different quantities and offers analytical solutions in some conditions generally\cite{18,19,20,21}. However, in the original definitions both the quantum discord and the geometric quantum discord are not symmetric with respect to the subsystems. For a symmetric extension of the quantum discord the global quantum discord has been presented in Ref.~\cite{22}. Furthermore, a geometric quantum discord for multipartite states, called geometric global quantum discord (GGQD), has been proposed in Ref.~\cite{23}. Nevertheless, similar to the original discord, it is extremely difficult to calculate the GGQD for generally given quantum states. In this paper, we study the GGQD for arbitrary two-qubit systems, and derive explicit expressions.

The paper is organized as follows. In section II  we review GQD and GGQD.
We derive an analytical formula of GGQD for arbitrary two-qubit states.
In section III, as examples we work out the GGQD for X-states.
Conclusions and discussions are given in section IV.

\section{Geometric Global Quantum Discord of two-qubit states}
For a bipartite state $\rho_{\mathfrak{AB}}$ in a composite system $\mathfrak{AB}$,
the total correlation between $\mathfrak{A}$ and $\mathfrak{B}$ is measured by the quantum mutual information
$$
I(\rho_{\mathfrak{AB}})=S(\rho_\mathfrak{A})-S(\rho_\mathfrak{A}|\rho_\mathfrak{B}),
$$
where $\rho_\mathfrak{A}$, $\rho_\mathfrak{B}$ are the
reduced density matrices associated with the subsystems $\mathfrak{A}$ and $\mathfrak{B}$, $S(\rho_\mathfrak{A}|\rho_\mathfrak{B})$ is
conditional entropy, $S(\rho)$$=-$Tr$(\rho\log_2\rho)$ is the von Neuman entropy.
One may also get the following quantity to characterize the quantum mutual information,
$$
J(\rho_{\mathfrak{AB}})=S(\rho_\mathfrak{A})-S(\rho_{\mathfrak{AB}}|\{\Pi_\mathfrak{B}^j\}),
$$
where $S(\rho_{\mathfrak{AB}}|\{\prod_\mathfrak{B}^j\})=\sum_{j}p_{j}S(\rho_{\mathfrak{A}|j})$,
$\rho_{\mathfrak{A}|j}=\frac{1}{p_j}\langle b_j|\rho_{\mathfrak{AB}}|b_j\rangle $, $\{\prod_\mathfrak{B}^j= |b_j\rangle\langle b_j|\}$ is a set of projectors,
$p_j$ denotes the probability of obtaining the $j$th measurement outcome.

The quantities $I(\rho_{\mathfrak{AB}})$ and $J(\rho_{\mathfrak{AB}})$ are equal in the classical case. However they are differnt in the quantum case.
The difference defined by $D(\rho_{\mathfrak{AB}})=I(\rho_{\mathfrak{AB}})-J(\rho_{\mathfrak{AB}})$ is called the quantum discord of the $\rho_{\mathfrak{AB}}$.
As the measurement is single side measurement of bipartite system, the global quantum discord $D(\rho_{\mathfrak{A}_1\mathfrak{A}_2\cdots \mathfrak{A}_N})$
for an arbitrary multipartite state $\rho_{\mathfrak{A}_1\mathfrak{A}_2\cdots \mathfrak{A}_N}$ is defined by
$$
D(\rho_{\mathfrak{A}_1\mathfrak{A}_2\cdots \mathfrak{A}_N}) = \min\limits_{\{\Pi_k\}}[S(\rho_{\mathfrak{A}_1\mathfrak{A}_2\cdots \mathfrak{A}_N}|\Phi(\rho_{\mathfrak{A}_1\mathfrak{A}_2\cdots \mathfrak{A}_N}))-\sum\limits_{j=1}\limits^{N}S(\rho_{\mathfrak{A}_j}|\Phi_j(\rho_{\mathfrak{A}_j}))],
$$
under all local measurements $\{\Pi_{\mathfrak{A}_1}^{j_1}\otimes\cdots\otimes\Pi_{\mathfrak{A}_N}^{j_N}\}$,
where $\Phi_j(\rho_{\mathfrak{A}_j}) = \sum\limits_{i}\Pi_{\mathfrak{A}_i}^{i}\rho_{\mathfrak{A}_j}\Pi_{\mathfrak{A}_i}^{i}$ and
$\Phi(\rho_{\mathfrak{A}_1\mathfrak{A}_2\cdots \mathfrak{A}_N}) = \sum\limits_k \Pi_k\rho_{\mathfrak{A}_1\mathfrak{A}_2\cdots \mathfrak{A}_N}\Pi_k $,
with $\Pi_k = \Pi_{\mathfrak{A}_1}^{j_1}\otimes\cdots\otimes\Pi_{\mathfrak{A}_N}^{j_N}$ and $k$ denoting the index string $(j_1\cdots j_N)$.

Following the concept of global quantum discord, the geometric global quantum discord (GGQD) is defined by
\begin{equation*}
D^{GG}(\rho_{\mathfrak{A}_1\mathfrak{A}_2\cdots \mathfrak{A}_N}) = \min\limits_{\sigma_{\mathfrak{A}_1\mathfrak{A}_2\cdots \mathfrak{A}_N}}\{\mathrm{Tr}[\rho_{\mathfrak{A}_1\mathfrak{A}_2\cdots \mathfrak{A}_N}-\sigma_{\mathfrak{A}_1\mathfrak{A}_2\cdots \mathfrak{A}_N}]^2 ~|~ D(\sigma_{\mathfrak{A}_1\mathfrak{A}_2\cdots \mathfrak{A}_N})=0\},
\end{equation*}
which is equivalent to the result in Ref.~\cite{23},
\begin{equation}\label{5}
D^{GG}(\rho_{\mathfrak{A}_1\mathfrak{A}_2\cdots \mathfrak{A}_N}) = \sum\limits_{\alpha_1,\alpha_2,\cdots,\alpha_N}C^2_{\alpha_1\alpha_2\cdots\alpha_N}-\max\limits_{\Pi}\sum\limits_{i_1i_2\cdots i_N}(\sum\limits_{\alpha_1,\alpha_2,\cdots,\alpha_N}A_{\alpha_1i_1}A_{\alpha_2i_2}\cdots A_{\alpha_Ni_N}C_{\alpha_1\alpha_2\cdots\alpha_N})^2,
\end{equation}
where $C_{\alpha_1\alpha_2\cdots\alpha_N}$ and $A_{\alpha_ki_k}$ are determined as follows.
For any $k$ ($1\leq k\leq N$), let $L(H_k)$ be the real Hilbert space consisting of all Hermitian operators on $H_k$, with the
inner product $\langle X|X^\prime\rangle = \mathrm{Tr}(XX^\prime)$ for $X$, $X^\prime\in L(H_k)$, for all $k$, and for given orthonormal basis
$\{ X_{\alpha_k}\}^{n^2_k}_{\alpha_k=1}$, $\{|i_k\rangle \}^{n_k}_{i_k=1}$ of $L(H_k)$ ,$H_k$.
$C_{\alpha_1\alpha_2\cdots\alpha_N}$ and $A_{\alpha_ki_k}$ are given by the following equations,
$$
\rho_{\mathfrak{A}_1\mathfrak{A}_2\cdots \mathfrak{A}_N}=\sum\limits_{\alpha_1,\alpha_2,\cdots,\alpha_N}C_{\alpha_1\alpha_2\cdots\alpha_N}X_{\alpha_1}\otimes X_{\alpha_2}\otimes\cdots\otimes X_{\alpha_N}
$$
and
$$
A_{\alpha_ki_k}=\langle i_{k}|X_{\alpha_{k}}|i_{k} \rangle.
$$

Now consider the GGQD of two-qubit states. For bipartite qubit states $\rho_{\mathfrak{AB}}$, Eq. (\ref{5}) can be simplified,
$$
D^{GG}(\rho_{\mathfrak{AB}}) = \sum\limits_{\alpha_1,\alpha_2}C^2_{\alpha_1\alpha_2}-\max\limits_{\Pi}
\sum\limits_{i_1i_2}(\sum\limits_{\alpha_1,\alpha_2}A_{\alpha_1i_1}A_{\alpha_2i_2}C_{\alpha_1\alpha_2})^2.
$$
Moreover, $\{X_m = \frac{\sigma_m^\mathfrak{A}}{\sqrt{2}} \}$, $\{Y_n = \frac{\sigma_n^\mathfrak{B}}{\sqrt{2}} \}$ are the
orthonormal bases, with $\sigma^\mathfrak{A}_m$, $\sigma^\mathfrak{B}_n$, $m,n=0,1,2,3$, are the Pauli matrices associated with the subsystems
$\mathfrak{A}$ and $\mathfrak{B}$, respectively. Therefore,
$$
D^{GG}(\rho_{\mathfrak{AB}})= \mathrm{Tr}(CC^\prime)-\max\limits_{AB}\mathrm{Tr}(ACB^\prime BC^\prime A^\prime),
$$
with $A = (A_{im})$, $B = (B_{jn})$, $A_{im} = \mathrm{Tr} (|i\rangle\langle i|X_m)$, $B_{jn} = \mathrm{Tr} (|j\rangle\langle j|Y_n)$,
where $\{|i\rangle \}$ and $\{|j\rangle \}$ are orthonormal bases. $C = (C_{mn})$ is given by $C_{mn} = \mathrm{Tr}\rho_{\mathfrak{AB}}X_m\otimes Y_n$.
From a similar approach in Ref.~\cite{16}, the matrices $C$, $A$ and $B$ can be written in the following forms,
\begin{equation}\label{10}
C= (C_{mn}) = \frac{1}{2}\left(\begin{array}{cc}
                            1 & y^\prime \\
                            x & T
                          \end{array}\right),
\end{equation}
$$
A = \frac{1}{\sqrt{2}}\left(\begin{array}{cc}
                            1 & a \\
                            1 & -a
                          \end{array}\right), ~~~~a = (a_1, a_2, a_3) = \sqrt{2}(A_{11}, A_{12}, A_{13}),
$$
$$
B = \frac{1}{\sqrt{2}}\left(\begin{array}{cc}
                            1 & b \\
                            1 & -b
                          \end{array}\right), ~~~~b = (b_1, b_2, b_3) = \sqrt{2}(B_{11}, B_{12}, B_{13})
$$
and
\begin{equation}\label{13}
\mathrm{Tr}(ACB^\prime BC^\prime A^\prime) = \frac{1}{4}[1+y^\prime b^\prime by+a(xx^\prime+Tb^\prime bT^\prime)a^\prime].
\end{equation}

Note that under local unitary transformations, any two-qubit state can be written as
$$
\rho_{\mathfrak{AB}} = \left(\begin{array}{cccc}
                                                                                    \rho_{00} & \rho_{01} & \rho_{02} & \rho_{03} \\
                                                                                    \rho_{01}^{\ast} & \rho_{11} & \rho_{12} & \rho_{13} \\
                                                                                    \rho_{02}^{\ast} & \rho_{12}^{\ast} & \rho_{22} & \rho_{23} \\
                                                                                    \rho_{03}^{\ast} & \rho_{13}^{\ast} & \rho_{23}^{\ast} & \rho_{33}
                                                                                  \end{array}
 \right).
$$
 Therefore
 \begin{align}\label{15}
 C &= \frac{1}{2}\left(\begin{array}{cccc}
                         \rho_{00}+\rho_{11}+\rho_{22}+\rho_{33} & 2(\rho_{01}+\rho_{23}) & 0 & \rho_{00}-\rho_{11}+\rho_{22}-\rho_{33} \\
                         2(\rho_{02}+\rho_{13}) & 2(\rho_{12}+\rho_{03}) & 0 & 2(\rho_{02}-\rho_{13}) \\
                         0 & 0 & 2(\rho_{12}-\rho_{03}) & 0 \\
                         \rho_{00}+\rho_{11}-\rho_{22}-\rho_{33} & 2(\rho_{01}-\rho_{23}) & 0 & \rho_{00}-\rho_{11}-\rho_{22}+\rho_{33}
                       \end{array}
  \right)\nonumber\\[3mm]
  &= \frac{1}{2}\left(\begin{array}{cccc}
      c_{00} & c_{01} & 0 & c_{03} \\
      c_{10} & c_{11} & 0 & c_{13} \\
      0 & 0 & c_{22} & 0 \\
      c_{30} & c_{31} & 0 & c_{33}
    \end{array}\right).
\end{align}
Then from Eq.(\ref{10}) we have
\begin{equation}\label{16}
x = \left(\begin{array}{c}
       2(\rho_{02}+\rho_{13}) \\[1mm]
       0 \\[1mm]
       \rho_{00}+\rho_{11}-\rho_{22}-\rho_{33}
     \end{array}\right),
\end{equation}
\begin{equation}\label{17}
y^\prime = \left(\begin{array}{ccc}
         2(\rho_{01}+\rho_{23}), & 0, & \rho_{00}-\rho_{11}+\rho_{22}-\rho_{33}
       \end{array}\right),
\end{equation}
\begin{equation}\label{18}
T = \left(\begin{array}{ccc}
             2(\rho_{12}+\rho_{03}) & 0 & 2(\rho_{02}-\rho_{13}) \\[1.5mm]
             0 & 2(\rho_{12}-\rho_{03}) & 0 \\[1.5mm]
             2(\rho_{01}-\rho_{23}) & 0 & \rho_{00}-\rho_{11}-\rho_{22}+\rho_{33}
           \end{array}
 \right).
\end{equation}
Substituting Eq.(\ref{16})-(\ref{18}) into Eq.(\ref{13}), we obtain
$$
\begin{aligned}
\mathrm{Tr}(ACB^\prime BC^\prime A^\prime) = &\frac{1}{4}[(c_{00}^2+c_{01}+c_{03}^2)+(c_{10}^2+c_{11}+c_{13}^2)a_1^2+(c_{30}^2+c_{31}+c_{33}^2)a_3^2+2(c_{10}c_{30}+c_{11}c_{31}+c_{13}c_{33})a_1a_3\\
+&2c_{01}c_{03}b_1b_3+2c_{01}c_{12}a_2b_1b_2+2c_{03}c_{22}a_2b_2b_3+c_{22}^2a_{2}^2b_{2}^2\\
+&2c_{11}c_{13}a_{1}^2b_1b_3+2c_{31}c_{33}a_3^2b_1b_3+2(c_{13}c_{33}+c_{13}c_{31})a_1a_3b_1b_3].
\end{aligned}
$$\\
The key point in calculating GGQD is to obtain the maximal value of $\mathrm{Tr}(ACB^\prime BC^\prime A^\prime)$. Let
\begin{equation}\label{20}
\begin{aligned}
f = &(c_{00}^2+c_{01}+c_{03}^2)+(c_{10}^2+c_{11}+c_{13}^2)a_1^2+(c_{30}^2+c_{31}+c_{33}^2)a_3^2+2(c_{10}c_{30}+c_{11}c_{31}+c_{13}c_{33})a_1a_3\\
+&2c_{01}c_{03}b_1b_3+2c_{01}c_{12}a_2b_1b_2+2c_{03}c_{22}a_2b_2b_3+c_{22}^2a_{2}^2b_{2}^2\\
+&2c_{11}c_{13}a_{1}^2b_1b_3+2c_{31}c_{33}a_3^2b_1b_3+2(c_{13}c_{33}+c_{13}c_{31})a_1a_3b_1b_3.
\end{aligned}
\end{equation}
Set $M_0 = (c_{00}^2+c_{01}+c_{03}^2)+(c_{10}^2+c_{11}+c_{13}^2)a_1^2+(c_{30}^2+c_{31}+c_{33}^2)a_3^2+2(c_{10}c_{30}+c_{11}c_{31}+c_{13}c_{33})a_1a_3$,
$M_{13} = 2c_{01}c_{03}+2c_{11}c_{13}a_1^2+2c_{31}c_{33}a_3^2+2(c_{11}c_{33}+c_{13}c_{31})a_1a_3$,
$M_{12} = 2c_{01}c_{22}a_2$,
$M_{23} = 2c_{03}c_{22}a_2$ and $M_{22} = c_{22}^2a_{2}^2$.
Then $f = M_0+M_{13}b_1b_3+M_{12}b_1b_2+M_{23}b_2b_3+M_{22}b_2^2$.
To obtain the maximal value of $\mathrm{Tr}(ACB^\prime BC^\prime A^\prime)$ we just need to obtain the maximal value of $\displaystyle\frac{1}{4}f$.

By taking a coordinate transformation $b_1 = \cos\theta_1\sin\theta_2$, $b_2 = \sin\theta_1\sin\theta_2$ and $b_3 = \cos\theta_2$, we have
\begin{numcases}{}
\frac{\partial f}{\partial\theta_1} = -M_{13}\sin\theta_2\cos\theta_2\sin\theta_1+M_{23}\sin\theta_2\cos\theta_2\cos\theta_1-M_{12}
\sin\theta_2\cos\theta_2\sin\theta_1+M_{22}\sin^2\theta_2\sin\theta_1\cos\theta_1 = 0 \notag, \\
\frac{\partial f}{\partial\theta_2} =M_{13}\cos\theta_1\cos^2\theta_2-M_{13}\cos\theta_1
\sin^2\theta_2+M_{23}\sin\theta_1\cos^2\theta_2-M_{23}\sin\theta_1\sin^2\theta_2 \notag \\
~~~~~~~~+M_{12}\cos\theta_1\cos^2\theta_2-M_{12}\cos\theta_1\sin^2\theta_2+2M_{22}\sin^2\theta_1\sin\theta_2\cos\theta_2 = 0 \notag.
\end{numcases}
The solutions of the above two equations can be divided into the following twelve cases:
\begin{enumerate}
  \item  $\cos^2\theta_1=(M_{13}-M_{23}+M_{12})^2$,\\
  $\cos^2\theta_2=\displaystyle\frac{M_{22}\sin^2\theta_1+\sqrt{(M_{13}
  \cos\theta_1+M_{23}\sin\theta_1+M_{12}\cos\theta_1)^2+M_{22}^2\sin^4\theta_1}}
  {2\sqrt{(M_{13}\cos\theta_1+M_{23}\sin\theta_1+M_{12}\cos\theta_1)^2+M_{22}^2\sin^4\theta_1}}$;
  \item  $4M_{22}^2(M_{13}-M_{23}+M_{12})\cos\theta_1-4M_{22}^2(M_{13}-M_{23}+M_{12})\cos^3\theta_1-4M_{22}^2(M_{13}+M_{23})\cos^3\theta_1
  +(M_{13}-M_{23}+M_{12})^2(M_{13}+M_{12})\cos\theta_1-4M_{22}^2M_{23}\cos^2\theta_1\sin\theta_1+(M_{13}-M_{23}+M_{12})^2M_{23}\sin\theta_1=0$.

  \item  $\cos^2\theta_1=\frac{M^2_{23}}{(M_{12}+M_{13})^2+M_{23}^2}$, $\sin^2\theta_1=\frac{(M_{12}+M_{23})^2}{(M_{12}+M_{13})^2+M_{23}^2}$, $\theta_2=\{0, \pi\}$;
  \item  $\theta_1=\{0, \pi\}$, $\theta_2=\{0,\pi\}$, $M_{13}+M_{12}=0$;
  \item  $\theta_1=\{0, \pi\}$, $\theta_2=\{\frac{\pi}{4}, \frac{3\pi}{4}\}$, $M_{23}=0$;
  \end{enumerate}
Substituting the above solutions of $\frac{\partial f}{\partial \theta_1}=\frac{\partial f}{\partial \theta_2}=0$ into Eq.(\ref{20}), $f$ becomes a function of the parameters $a_1$, $a_2$ and $a_3$. Further setting $a_1=\cos\theta_3\sin\theta_4$, $a_2=\sin\theta_3\sin\theta_4$, $a_3=\cos\theta_4$ in $\max\limits_{\theta_1,\theta_2}f$, we can repeat the above procedure to find $\max\limits_{A,B}\mathrm{Tr}(ACB^\prime BC^ \prime A^\prime)=\displaystyle\frac{1}{4}\max\limits_{\theta_1,\theta_2,\theta_3,\theta_4}f=\frac{1}{4}\max\limits_{\theta_3,\theta_4}\max\limits_{\theta_1,\theta_2}f$.
Here the value of $\max\limits_{\theta_1,\theta_2}f$ depends on $M_{ij}$ which is a function of $\theta_3$ and $\theta_4$.

As we know, it is too difficult to calculate the exact value of geometric global quantum discord\cite{23}. Nevertheless, our method above can calculate it and some detailed examples will be given in the
next section.

\section{Examples For Geometric Global Quantum Discord}
We now apply our approach to calculate some two-qubit states. Let us first consider X-states\cite{24}, which, under local unitary transformations, have the form
\begin{equation}\label{21}
\rho_{\mathfrak{AB}} = \left(\begin{array}{cccc}
                                                                                    \rho_{00} & \rho_{01} & \rho_{02} & \rho_{03} \\
                                                                                    \rho_{01}^{\ast} & \rho_{11} & -\rho_{03} & \rho_{13} \\
                                                                                    \rho_{02}^{\ast} & -\rho_{03}^{\ast} & \rho_{22} & \rho_{23} \\
                                                                                    \rho_{03}^{\ast} & \rho_{13}^{\ast} & \rho_{23}^{\ast} & \rho_{33}
                                                                                  \end{array}
 \right).
 \end{equation}
We have
\begin{equation}
f = (c_{00}^2+c_{01}^2)+(c_{10}^2+c_{13}^2)a_1^2+(c_{30}^2+c_{33}^2)a_3^2+2(c_{10}c_{30}+c_{13}c_{33})a_1a_3+2c_{01}c_{22}a_2b_1b_2+c_{22}^2a_{2}^2b_{2}^21b_3.
\end{equation}
Arranging from above solutions, we get $b_1=0$, $b_2=0$, $b_3=\{1, -1\}$
\begin{equation}\label{33}
\begin{aligned}
\max_{AB}f &= \max_{\theta_3,\theta_4}[(c_{00}^2+c_{01}^2)+(c_{10}^2+c_{13}^2)a_1^2+(c_{30}^2+c_{33}^2)a_3^2+2(c_{10}c_{30}+c_{13}c_{33})a_1a_3]\\
&=\max_{a_1,a_2,a_3}[(c_{00}^2+c_{01}^2)+(c_{10}^2+c_{13}^2)a_1^2+(c_{30}^2+c_{33}^2)a_3^2+2(c_{10}c_{30}+c_{13}c_{33})a_1a_3].
\end{aligned}
\end{equation}
 Since $a_1^2+a_{2}^2+a_{3}^2=1$ and $a_2$ does not appear in $f$, we set $a_2=0$ and $a_1=\cos\theta_3$, $a_1=\sin\theta_3$. Then
$$
f=(c_{00}^2+c_{01}^2+c_{10}^2+c_{13}^2)+(c_{30}^2+c_{33}^2-c_{10}^2-c_{13}^2)\sin^2\theta_3+2(c_{10}c_{30}+c_{13}c_{33})\sin\theta_3\cos\theta_3
$$
and
$$
\frac{\partial f}{\partial \theta_3}=(c_{30}^2+c_{33}^2-c_{10}^2-c_{13}^2)\sin2\theta_3+2(c_{10}c_{30}+c_{13}c_{33})\cos2\theta_3=0,
$$
which gives rise to either $\theta_3=\{\frac{\pi}{4},\frac{3\pi}{4}\}$ if $c_{30}^2+c_{33}^2-c_{10}^2-c_{13}^2=0$, or
$
\theta_3=\frac{1}{2}\arctan \frac{2(c_{10}c_{30}+c_{13}c_{33})}{c_{30}^2+c_{33}^2-c_{10}^2-c_{13}^2}
$
if $c_{30}^2+c_{33}^2-c_{10}^2-c_{13}^2\neq 0$. Substituting the results into Eq.(11), we can obtain the GGQD of $\rho_{\mathfrak{AB}}$.

Now, we would like to show a more detailed example, let us consider
\begin{eqnarray}
\rho =&\displaystyle{\frac{1}{4}}(I\otimes I-\sigma_y \otimes \sigma_y+C_3\sigma_z \otimes \sigma_z)\\& \nonumber \\
=&\left(\begin{array}{cccc}
                                                      1+C_3 & 0 & 0 & 1 \\
                                                      0 & 1-C_3 & -1 & 0 \\
                                                      0 & -1 & 1-C_3 & 0 \\
                                                      1 & 0 & 0 & 1+C_3
                                                    \end{array}\right),
\end{eqnarray}
which is a state of the form (\ref{21}). From Eq.(4) we can obtain
$$
C=\displaystyle{\frac{1}{2}}\left(\begin{array}{cccc}
                                                      1 & 0 & 0 & 0 \\
                                                      0 & 0 & 0 & 0 \\
                                                      0 & 0 & -1 & 0 \\
                                                      0 & 0 & 0 & C_3
                                                    \end{array}\right),
$$\\
furthermore $f=1+C_3^2a_3^2+a_2^2b_2^2=2+(C_3^2-1)a_3^2$. Furthermore, if $C_3^2-1\geq 0$, then $\max f=C_3^2+1$, hence $\max\mathrm{Tr}(ACB^\prime BC^\prime A^\prime)$ = $\frac{1}{4}$$(C_3^2+1)$, $\mathrm{Tr}(CC^\prime) = \frac{1}{4}(C_3^2+2).$
We have
\begin{equation}
D^{GG}(\rho)= \mathrm{Tr}(CC^\prime)-\max\limits_{AB}\mathrm{Tr}(ACB^\prime BC^\prime A^\prime)=\frac{1}{4},
\end{equation}
otherwise, if $C_3^2-1< 0$, then $\max f=2$, $\max\mathrm{Tr}(ACB^\prime BC^\prime A^\prime) = \frac{1}{2}$, $\mathrm{Tr}(CC^\prime)=\frac{1}{4}(C_3^2+2)$.\\
We have
\begin{equation}
D^{GG}(\rho)= \mathrm{Tr}(CC^\prime)-\max\limits_{AB}\mathrm{Tr}(ACB^\prime BC^\prime A^\prime)=\frac{1}{4}C_3^2.
\end{equation}
In conclusion, Eq.(12) and Eq.(13) can be written in a uniformed equation
$$
D^{GG}(\rho)= \mathrm{Tr}(CC^\prime)-\max\limits_{AB}\mathrm{Tr}(ACB^\prime BC^\prime A^\prime)=\frac{1+C_3^2-\max\{ 1,C_3^2\}}{4}.
$$
It is remarked that Example 2. in Ref.~\cite{23} has the same result with our example.

\section{Conclusions and discussions}

We have calculated the geometric global quantum discord for arbitrary two-qubit states. Although the geometric global quantum discord is controlled by many parameters of the quantum states, we analyze the symmetry of geometric global quantum discord and simplify the problem. Then we adopt our method to demonstrate how the parameter of two-qubit states influences the outcome. Furthermore, continuing our idea we work out the extremum problem which lies at the core of calculating the geometric global quantum discord for arbitrary two-qubit states and obtain the accurate solution of the geometric global quantum discord for arbitrary two-qubit states. Some detailed examples are also presented.


\end{document}